\documentclass[a4paper,10pt]{scrartcl}

\usepackage{amsmath,amsfonts,amssymb}
\usepackage{url}
\usepackage{fullpage}
\usepackage{hyperref}
\usepackage{graphicx}

\newcommand{\CI}{\mathbb{CI}}
\newcommand{\ACI}{\mathbb{ACI}}
\newcommand{\bbbn}{\mathbb{N}}

\newcommand{\component}[1]{\langle #1 \rangle}

\newcommand{\qqop}[1]{\mathrel{\makebox[2em]{$#1$}}}

\newcommand{\qqeq}{\qqop{=}}

\newtheorem{proposition}{Proposition}
\newtheorem{definition}{Definition}

\title{A Model for Configuration Management of Open Software Systems\thanks{This work was partially funded by Regione Toscana as part of the research activities of the T-OSSLab project}}
\author{\phantom{SPAZIO}\\ 
Roberto Barbuti \qquad Giovanni A. Cignoni \qquad Paolo Milazzo\\
\phantom{SPAZIO}\\
\\
\large
Dipartimento di Informatica, Universit\`a di Pisa\\
\large
Largo Pontecorvo 3, 56127 Pisa, Italy.\\
\large
\texttt{\{barbuti,cignoni,milazzo\}@di.unipi.it}\\
\phantom{SPAZIO}\\
}

\begin{document}
\maketitle

\begin{abstract}
The article proposes a model for the configuration
management of open systems. The model aims at validation of configurations
against given specifications. An extension of decision graphs is proposed to
express specifications. The proposed model can be used by software
developers to validate their own configurations across different versions of the
components, or to validate configurations that include components by third
parties. The model can also be used by end-users to validate compatibility among
different configurations of the same application. The proposed model is first
discussed in some application scenarios and then formally defined. Moreover,
a type discipline is given to formally define validation of a configuration
against a system specification.
\end{abstract}

\section{Introduction}

Defining a software product as a system of components is common both in theory
and practice of Software Engineering \cite{TOSEM05}. Many models and standards 
\cite{ISO12207,IEEE828} of 
the software development process consider {\em configuration management} as the 
set of activities for identifying, controlling and managing software components 
during the life-cycle of the product, in particular at project milestones when 
baselines and releases have to be arranged and frozen. 

The wide availability of open formats and the large success of open source
projects make configuration management a constantly growing issue. Many
developers have the possibility to create add-ons or even to modify the core
components of the application to originate a different one. 
On the user side, Internet distribution allows everyone to choose among
available
components and to download them to extend and update the applications. 

As a result, when open systems are involved, identification of components,
selection 
of their right version, control over their composition, i.e. configuration
management,
does no longer belong to the closed environment of a single software house, but
to
a wide context participated by many developers and where the end user is an
active agent (seriously concerned by the reliability of the results).

In this paper we propose a model for the configuration management of open system.
The model
aims at the validation of configurations with respect to given specifications.
Specifications may be defined by the original software developer to validate
its 
own configurations across different versions of the components or to validate
configurations that include components by third parties. Specifications may
even 
be defined by users themselves to validate compatibility among different 
configurations of the same application.

The proposed model is first discussed in some application scenarios and then formally defined. Moreover,
a type discipline \cite{Card96} is given to formally define validation of a configuration
against a system specification.

\section{Scenarios and Basic Needs}
There are many examples of applications in which a end user is involved in
configuration management. The scenarios presented in the following help to
identify
different needs and to define the basic operations of configuration
management
that our model aims to support.

\subsection{Scenarios}\label{sec:scenarios}

Internet browsers are a simple and common example of end user configuration 
management. Browsers are characterized by frequent updates, often needed for 
security reasons, and by many plug-ins for media visualization. 
The large majority of users passively perform the installation of all suggested 
updates and add-ons, but sometimes someone has (or wants) to decide what she/he 
likes to install.

A more complex and interesting example is given by software development
environments.
Such tools usually offer support for different compilers, libraries, debuggers,
editors,
modelling and code generation tools, and so on. {\em Eclipse} and {\em Cygwin} are
two 
good examples, but there are many others. 
In these cases, the user is also concerned 
by {\em compatibility} among configurations in a given context. For instance, a
programmer
wants to install a new add-on to experiment with a new UML modeling tool, but
also 
needs to check compatibility of the wished upgrade with all the installations of
the 
colleagues working in the same software project.

On a greater scale, Linux distributions are probably the most complex case of
large
software system configuration management. A distribution offers a kernel
version, a 
number of device drivers, a choice of system tools, several software utilities, 
and a number of applications. All of these software artifacts have dependencies
that
must be respected. The user willing to install 
a new application on top of the basic distribution is involved in a process that may lead to installing other
artifacts,
like libraries, data sets and so on, that can conflict or be incompatible with
the 
previously installed ones. The usual solution is to provide official
repositories 
which collect certified compatible versions of the most common applications
(repositories 
are valuable assets in the success of distributions like {\em Ubuntu}, {\em Fedora} and the
others).
However, repositories are never complete and usually they apply a very simple
policy: 
provide just the last stable versions. This approach does not support the needs
of 
local groups of users, like software developers, or in general smart users of
applications. Such people need to define and control custom configurations of 
their working environments. 

An other interesting example is given by PC games. The typical game architecture
is based 
on a game engine able to interpret a number of contents (maps, 3D models, AI
scripts, ...).
The engine may be closed, but, even for proprietary games, contents often have
open
formats and third parties, as well as users themselves, may develop new {\em
mods}, namely modifications of contents that provide new levels, new game situations and so
on. 
Players are often organized in communities, less or more independent from the
game 
developer, that build and share mods of their favorite game \cite{EWSPT01}. 
Technically speaking, a mod is a configuration made from some parts of the
original 
game and a number of new ones. 
In this scenario, configuration control is not only a matter of reliability
(the 
game has to run) but also a requirement to guarantee fair play. For instance, an
easy
way for a first person shooter player to cheat is to mod its local configuration
to 
give brilliant textures to the avatars of the other network players. This is
another 
case where checking the compatibility among configurations is needed. 

\subsection{Basic Operations}

There are many examples of automated procedures to install, configure and update
the applications. Configuration control of commercial closed applications has
the 
advantage of complete control both of the software architecture and of the 
development process.
In contexts where applications are open, the number of agents involved in
software
development and in the management of its configurations largely increases: a
shared 
model to describe and to manage configurations is needed. 

In common speech there is not difference in the usage of terms like a
``update", 
``extension", ``version" and so on. Our model describes such operations in terms
of 
operations on {\em components} (that, at the end, are files or set of files).
The model has three basic operations:

\begin{itemize}
\item {\em update}, that is the substitution of components of the current
configuration 
with more recent versions of them;
\item {\em extension}, that is the addition to the current configuration of new
components
aimed to increase the functionality of the software system;
\item {\em compatibility check}, that is the examination of a configuration in
order to verify a specification derived by another configuration.
\end{itemize}

An installed {\em configuration} is just a set of components. Update and
extension are 
operations on the set of components. In practice, such operations are easily
recognizable
in the previously introduced scenarios: a plug-in for movie reproduction can,
at 
some time, be added (extension) to a web browser and, some time later,
substituted 
with a new version (update); the same happens for a library in a development 
environment or a dungeon map in a game. 

Update and extension operations modify the installed configuration. The life of
the 
installation of a software system evolves in steps that always move from a
consistent
configuration to another consistent one. Of course, update and extension must 
be reversible.
Consistency is obtained by the respect of a set of constraints, i.e the {\em
configuration
specification}. The specification is defined by the architecture designer, but
it must 
leave enough freedom to allow the variability requested by other developers 
and end users. 

Compatibility check is an operation usually not identified as part of
configuration management (often not identified at all). The most similar
implementation
of the concepts are the built-in {\em anticheat} features implemented in many 
proprietary game platforms (like, for instance, {\em Steam}), but these are usually very restrictive and tied to the
official 
game configurations.
Scenarios where great openess and strict compatibility control must cohesist,
like 
the software development team, are uncovered by these kind of solutions.

For compatibility check, additional constraints are defined by users which want
to 
define the set of the configurations that fulfills their specific needs, that
generally,
is a subset of the whole set of consistent configurations.
The best way of defining such a set is by deriving the specification from a target
configuration. For instance, in a software development team the target
compatibility
configuration is the baseline tool-set or, in the network game scenario, the
target
compatibility configuration of the master player that sets the rules of the
{\em deathmatch}.

\section{Specification and Representation of Configurations}\label{sec:config}

In general, a software system, for instance one of the applications we use
everyday,
is perceived as something identified and unique. In practice a software system 
is a set of sets. This is the set of all working variations that can 
be built by combining the various versions of the components that belong to the 
software system or, more in general, to its runtime environment.

\bigskip

\noindent {\bf Software system}. 
A set of configurations, each one of them is a set of interacting components.

\bigskip

Intuitively, a component is a unit that may be separately distributed and
installed 
by the end user \cite{JSS01}.
In practice, a component is made by one or more files like
binaries, libraries, data sets in various formats, and so on. In our model,
the concept of {\em component type} is introduced to define the characteristics 
of a component of a software system. 
At the maximum level of detail, each file must be considered as a component. 
Our model permits several levels of details thus complying with those frequent
situations where, for practical reasons, it is convenient to see components as
set 
of several files that are always distributed togheter.

\bigskip

\noindent {\bf Configuration}. 
A set of components that complies to the {\em configuration specification} 
of the software system.

\bigskip

Specification of configurations of software systems is a well known problem. 
A classical approach uses decisions graphs to represent the possible choices in
the
building of a configuration starting from the set of components. The best known
example is given by {\em and/or graphs} \cite{CSUR98}.
Our model extends this approach to systems where the set of components is not
strictly defined from the beginning. This complies with open development, where 
the evolution of the set of components cannot be strictly controlled. 

The model introduces {\em configuration specification graphs} to specify and
represent the set of valid components types for a software system and their
composition rules to build the valid configurations of the software system.

\bigskip

\noindent {\bf Component Specification Node} (CSN). A component type that, 
in a configuration, can be instantiated into one or more components. 
Every CSN contains information on the described components, such as their names (nm), origins (org) and version numbers (ver).
Constraints can be imposed on such information such as restrictions (e.g. the name must start with a certain prefix p -- denoted p*) or consistency requirements (e.g. two componets must have the same version number).
Every CSN 
(leafs excluded) has an associated interval that bounds the number of
components 
that can instantiate the set of its CSN successors. The root node is the
component 
type that corresponds to the complete software system that can be instantiated 
into one of its valid configurations.

\bigskip

\noindent {\bf Composition Specification Arc} (CSA). A directed arc that
connects a 
CSN to its CSN successors that specify the types that may instantiate it in a
valid
configuration of the software system. Every CSA has an associated interval that 
bounds the number of components that may instantiate the pointed CSN sucessor.
For every interval $[ a, b ]$ associated to a CSN and for all intervals $[ a_i ,
b_i ]$
associated to its $i$ successor CSAs, it holds $\sum a_i \le a \le b \le \sum
b_i$.

\bigskip

\noindent {\bf Dependence Specification Arc} (DSA) a directed arc that
connects 
a CSN to an other CSN and expresses the possibility that, in a configuration,
an 
instance of the former need the presence of an instance of the latter.

\bigskip

\noindent {\bf Configuration Specification Graph}. 
A directed graph with a set of CSNs, a set of CSAs and a set of DSAs. If only
CSNs and CSAs are considered, the CSG is a tree.

\bigskip

\begin{figure}
\centering
\includegraphics[width=11cm]{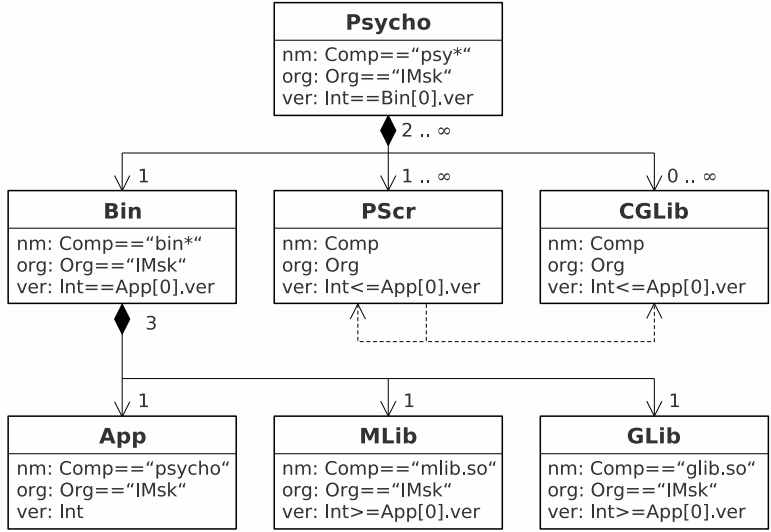}
\caption{A Configuration Specification Graph for a simple application}\label{fig:csg}
\end{figure}

\autoref{fig:csg} shows the CSG of a very simple software system. The used nototion is 
borrowed from UML \cite{UML2.3} {\em class diagrams}.
The system is {\em Psycho}, an hypothetical application made by the {\em IMsk}
software 
company that reproduces audio files while visualizing programmable psychedelic
effects
on the video. 
To produce the effects the application interprets a simple script language that
calls
graphic primitives from libraries built under respect of given APIs. 
For smart users, a large part of the fun in using Psycho is programming 
scripts and sharing them with friends. Geeks enjoy coding new libraries
featuring
more sophisticated graphic primitives.

The root CSN identifies the software system, Psycho, as a type that can be 
instantiated in one of its valid configurations.
Each Psycho configuration is made by two or more components. 
The leftmost component type {\em Bin}, which is the core part of the
application, 
is made by exactly three components types:
{\em App}, {\em MLib} and {\em GLib}.
Each of these component types, in this example, can be instantiated by just one 
component. We can imagine them as actual files: an executable and two dynamic
libraries,
one for audio files reproduction and the other for the default set of graphic
effects.
For the Bin component this CSG is quite strict: the intervals associated to CSNs
and CSAs
imply a fixed number of components, the specification of attributes implies
given
identifiers and there are constraints on version numbers and origin of the
components.

In particular, the version number of libraries must be greater than the version 
of the main executable, as IMsk allows patches of libraries but, if the core 
application is modified, it must be released with aligned versions of libraries.
Moreover,  this CSG implies that the Bin components have to be released only by
IMsk, 
as the origin constraint specifies. In this example to specify the origin we
used 
simple labels, however, in the implementation of the model origin is better
specifid 
by unique identifiers, as for instance URLs, that both provide uniqueness and
ability
to directly refer a distribution repository.

These strict rules apply only to the Psycho software system. Depending on the
software
license, it will be possible for others to define a new CSG with weaker
constraints.
It could specify another software system that is not Psycho, albeit tied to the
original
Psycho and sharing with it many components. This is just the kind of control
that open
source projects need when they approach the twilight zone where configurations
mix 
with project forks.

On its right side, the Psycho CSG is more open to extensions: scripts and custom
graphic
libraries can added by everyone. The only constraints are that at least one
script 
must exists (e.g. the one shipped with the original distribution by IMsk) and
that version numbers of scripts and custom libraries must be equal or lower
than 
the version number of the application. 

\begin{figure}
\centering
\includegraphics[width=9cm]{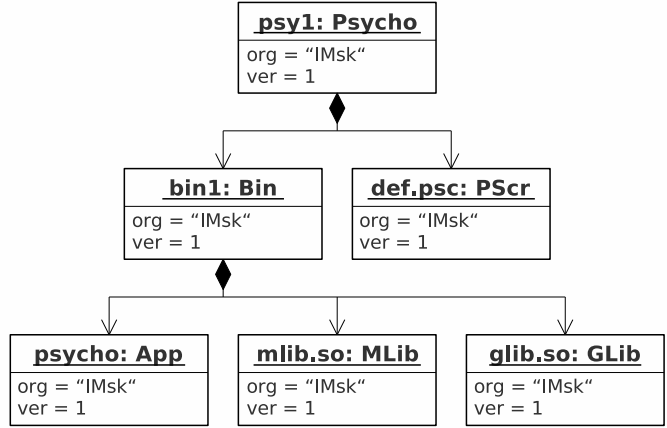}
\caption{A configuration compliant with the CSG of \autoref{fig:csg}.}\label{fig:ex1}
\end{figure}

\begin{figure}
\centering
\includegraphics[width=12cm]{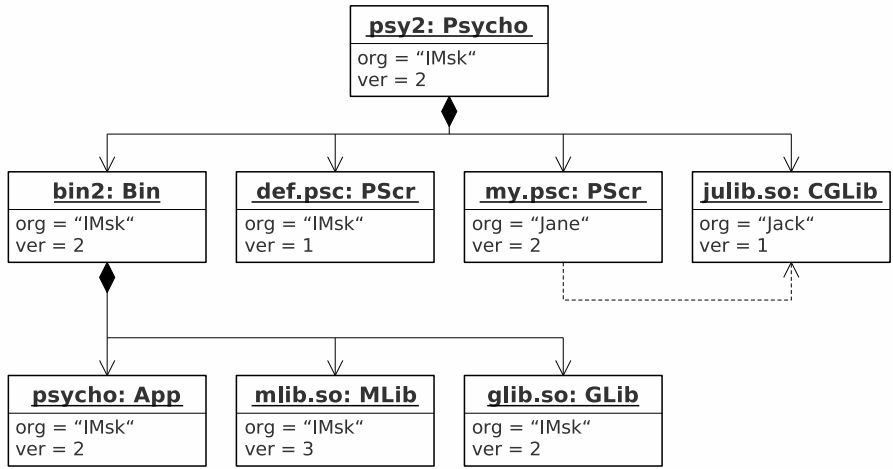}
\caption{Another configuration compliant with the CSG of \autoref{fig:csg} and compatible
with the configuration of \autoref{fig:ex1} (not the conversely).}\label{fig:ex2}
\end{figure}

In \autoref{fig:csg} are shown, as dotted arrows, two DSAs. The first is between PScr
and 
CGLib, as scripts may depend on graphic primitives supplied by custom
libraries; 
the second DSA is between PScr and PScr itself, as the script language allows
script
calls. DSAs are useful to express dependencies in the variable part of a
configuration.
Of course scripts depend on the application, but being the Bin subtree already 
specified as mandatory in all of its components, it is not necessary to express 
dependencies. Note that the CSG simply allows the possibility for such
dependences,
they actually depend on configurations (see \autoref{fig:ex1} and \autoref{fig:ex2}).

A CSG specifies the valid configurations of a software system. A single
configuration 
is described by a {\em Configuration Graph} (CG). \autoref{fig:ex1} and \autoref{fig:ex2} show two
examples of
CGs that describe two different configurations of the Psycho software system. 
Again, the notation is borrowed from UML: a CG is an {\em object diagram}. 
CGs and CSGs have a similar structure, but CG nodes are objects, instances of 
the CSG nodes, which are classes. Moreover, all composition relations in a CG
are
{\em 1 to 1}: where in a CSG there is a single node that can be instantiated in 
a variable number of components, in the CG all the components are represented.

The CG of \autoref{fig:ex1} represents a very basic installation of Psycho, that is
referred
as {\em psy1}. It has only the default script (def.psc) and no additional 
graphic library. All components are originated by IMsk and are marked as version
1.

The use of identifiers in CGs is worth a note. They are introduced to easily
refer
a CG, a subtree of a CG (like {\em bin1} in {\em psy1}) or a node of a CG.
However, it is important to higlight that identifiers of configurations are 
the CG themselves. 
More in details: for each node that is root of a CG subtree, its actual
identifier 
is the subtree.
The identifiers of nodes that are leaves can be easily and usefully associated
to
file names, but the version number is also needed to actually identify an 
installed leaf component.

The CG of \autoref{fig:ex2} shows an upgraded and customized installation: {\em psy2}. 
Versions are more recent and a new script is added (my.psc). 
Because such script needs a custom graphical library (julib.so, that provides 
functions to generate shapes inspired by the Julia set), the library
is installed too. Thus, the DSA between my.psc and julib.so represents a true 
dependence between two actual components - a dependence that was ``foreseen" 
in the CSG.

Besides the UML arc notation, the dependence information belongs to the
dependent 
component: the description of my.psc tells the installer about the
dependence and, if julib.so is not already present, it calls its installation. 
The same information can be used to prevent disinstallation of 
julib.so if my.psc is still present in the configuration.

The two CGs of \autoref{fig:ex1} and \autoref{fig:ex2} show also an example of compatibility among
configurations. Being {\em psy2} an updated superset of {\em psy1}, it is 
compatible with {\em psy1}, that is the user of {\em psy2} can enjoy all 
the psychedelic effects who see the user of {\em psy1}.
However, {\em psy1} is not compatible with {\em psy2}, that is the user of
{\em psy1} can't see the Julia effects.

\section{Formal Definition}

In this section we give a formal definition of configurations and of configuration specifications, followed by a formalization (by means of a type discipline -- see \cite{Card96} for a survey on type systems) of the procedure for checking the compliance of a configuration with respect to a given specification.

\subsection{Configurations and specifications}

A software component can be uniquely identified by its name, origin (i.e.
manufacturer) and version number. Moreover, the type associated with the
component in a software system describes unambiguously the role of such
a component in the system itself.

Let $Types$ be the finite set of all the types of component of a software
system. Moreover, let $Names$, $Origins$ and $Versions$ be the (possibly inifinite)
sets of all possible software component names, origins and versions,
respectively.

\begin{definition}[Component identifier] A {\em component identifier} is a tuple
$(t,n,o,v)$, where $ t \in Types, n \in Names, o \in Origins$ and $v \in
Versions$. We denote with $\CI$ the set of all component identifiers, namely
$\CI = Types \times Names \times Origins \times Versions$.
\end{definition}

A component identifier refers unanbiguously to a component of a specific software
system. In many cases, however, we shall need to refer to a set of components
of the same type. For instance, a Component Specification Node, namely a node of
a Configuration Specification Graph, may actually be instantiated in different configurations
with different components of the same type. In order to describe a CSN we shall need
an identifier (called abstract component identifier) for sets of components of the same type.

\begin{definition}[Abstract component identifier] An {\em abstract component
identifier} is a tuple $(t,N,O,V)$, where $t \in Types, N \subseteq Names, O
\subseteq Origins$ and $V \subseteq Versions$. We denote with $\ACI$ the set
of all abstract component identifiers, namely $\ACI = Types
\times \wp(Names) \times \wp(Origins) \times \wp(Versions)$.
\end{definition}

Let $Elements$ be the (possibly infinite) set of basic constituent elements
that may be included the software architecture (e.g. files). We now formally define
the notion of component and of configuration of a software system.

\begin{definition}[Component] A {\em component} is a tuple having one of the
following two forms:
\begin{itemize}
 \item $\component{ci,D,E}$ with $E \subseteq Elements$ and finite;
\item $\component{ci,D,CI}$ with $CI \subseteq \CI$ and finite;
\end{itemize}
where $ci \in \CI$ is the identifier of the component, $D \subseteq \CI$ is the finite set of components upon which component $ci$
depends.
\end{definition}

Given a component $c$ such that either $c = c' = \component{ci,D,E}$ or
$c = c'' = \component{ci,D,CI}$, let $ci(c) = ci$, $dependencies(c) = D$, $children(c') =
\varnothing$ and $children(c'') = CI$. Moreover, given a set of components $C$,
let $cis(C)$ be the extension of $ci(c)$ to sets of component specifications
defined in the obvious way. 

Note that it is not reasonable in a software system to explicitly specify
a dependence of a component from the subcomponents constituting it (since it is an obvious dependence). Hence, for a given component $c$ we will assume $dependencies(c) \cap children(c) = \varnothing$.

\begin{definition}[Configuration] A {\em configuration} is a set $C$ of
components such that:
\begin{enumerate}
 \item $\forall c \in C$ it holds $children(c) \subseteq cis(C)$;
 \item $\forall c \in C$ it holds $dependencies(c) \subseteq cis(C)$;
 \item $\exists ! c \in C$ s.t. $\forall c' \in C$ it holds $ci(c) \not\in
children(c')$;
\item $\forall c \in C$ either $\not\exists c'$ s.t. $ci(c) \in children(c')$ or $\exists ! c'$ s.t. $ci(c) \in children(c')$
\end{enumerate}
\end{definition}

A configuration is a set that actually represents a tree. In fact, each component of a configuration
is either a leaf of the tree (if it has the form $\component{ci,D,E}$) or an intermediate node
(if it has the form $\component{ci,D,CI}$) with components in $CI$ as children.
The conditions in the definition of configurations ensure that a configuration is a well-defined tree, namely
that all nodes are present in the configuration and that there is only one root.

It is easy to see that this definition of configuration corresponds to the notion of configuration
proposed in \autoref{sec:config}, with the sets of dependencies representing Dependence Specification Arcs.
For instance, the configuration in \autoref{fig:ex1} can be represented as follows. We have six components represented by
the following component identifiers:
\begin{gather*}
 ci\_psy_1 = (Psycho, psy1, IMsk, 1) \\
 ci\_bin_1 = (Bin, bin1, IMsk, 1) \\
 ci\_def.psc_1 = (PScr, def.psc, IMsk, 1) \\
 ci\_psycho_1 = (App, psycho, IMsk, 1) \\
 ci\_mlib.so_1 = (MLib, mlib.so, IMsk, 1) \\
 ci\_glib.so_1 = (Glib, glib.so, IMsk, 1)
\end{gather*}
that are used in the following representation of the configuration:
\begin{align*}
 C_{psy1} = \{ \;\;
  & \component{ci\_psy_1,\varnothing,\{ci\_bin_1,ci\_def.psc_1\}},\\
  & \component{ci\_bin_1,\varnothing,\{ci\_psycho_1,ci\_mlib.so_1,ci\_glib.so_1\}},\\
  & \component{ci\_def.psc_1,\varnothing,def.psc},\\
  & \component{ci\_psycho_1,\varnothing, psycho},\\
  & \component{ci\_mlib.so_1,\varnothing,mlib.so},\\
  & \component{ci\_glib.so_1,\varnothing,glib.so}
\;\; \}
\end{align*}

Similarly, the configuration in \autoref{fig:ex2} can be represented as follows. We have eight components represented by
the following component identifiers:
\begin{gather*}
 ci\_psy_2 = (Psycho, psy2, IMsk, 2) \\
 ci\_bin_2 = (Bin, bin2, IMsk, 2) \\
 ci\_def.psc_2 = (PScr, def.psc, IMsk, 1) \\
 ci\_my.psc_2 = (PScr, my.psc, Jane, 2) \\
 ci\_julib.so_2 = (CGLib, julib.so, Jack, 1)\\
 ci\_psycho_2 = (App, psycho, IMsk, 2) \\
 ci\_mlib.so_2 = (MLib, mlib.so, IMsk, 3) \\
 ci\_glib.so_2 = (Glib, glib.so, IMsk, 2)
\end{gather*}
that are used in the following representation of the configuration:
\begin{align*}
 C_{psy2} = \{ \;\;
  & \component{ci\_psy_2,\varnothing,\{ci\_bin_2,ci\_def.psc_2,ci\_my.psc_2,ci\_julib.so_2\}},\\
  & \component{ci\_bin_2,\varnothing,\{ci\_psycho_2,ci\_mlib.so_2,ci\_glib.so_2\}},\\
  & \component{ci\_def.psc_2,\varnothing,def.psc},\\
  & \component{ci\_my.psc_2,\{ci\_julib.so_2\},my.psc},\\
  & \component{ci\_julib.so_2,\varnothing,julib.so},\\
  & \component{ci\_psycho_2,\varnothing, psycho},\\
  & \component{ci\_mlib.so_2,\varnothing,mlib.so},\\
  & \component{ci\_glib.so_2,\varnothing,glib.so}
\;\; \}
\end{align*}

Now we formally define the notion of configuration specification, that will corresponds to a CSG. For the sake of simplicity we shall not consider consistency constraints between different components of a specification.
In the definition we shall use possibly infinite intervals over $\bbbn$. We denote with $Intervals$ the set of all possible intervals, 
namely $Intervals = \bbbn \times (\bbbn\cup\{\infty\})$.

Let summation over $\bbbn\cup\{\infty\}$ extend the usual summation over $\bbbn$ as follows: $n + \infty = \infty + n = \infty + \infty = \infty$.
Moreover, we define the following operations on intervals: summation $[n_1,n_1'] \boxplus [n_1,n_2'] = [n_1+n_2,n_1'+n_2']$, and inclusion $[n_1,n_1'] \subseteq [n_2,n_2'] = n_1 \geq n_2 \wedge n_1' \leq n_2'$.

\begin{definition}[Component specification] A {\em component specification} is
a tuple $\component{aci,AD,ACI,[n,n']}$ where $aci$ is the abstract identifier
of the component specification, $AD \subseteq \ACI$ is the finite set of
components upon which components specified by $aci$ depend, $ACI \subseteq
\ACI \times Intervals$ is a finite set of sub-component specifications each quantified
by an interval and such that $\forall (aci_1,[n_1,n_1']),(aci_2,[n_2,n_2']) \in ACI$, it
holds $aci_1 \neq aci_2$, and $[n_i,n_i'] \in Intervals$.
\end{definition}

\noindent Given a component specification $cs = \component{(t,N,O,V),AD,ACI,[n,n']}$, let
$aci(cs) = (t,N,O,V)$, $type(cs) = t$, $dependencies(cs) = AD$, and $children(cs) = \{aci \mid
(aci,[n_1,n_1']) \in ACI\}$. Moreover, given a set of component specifications $CS$,
let $acis(CS)$ be the extension of $aci(cs)$ to sets of component
specifications defined in the obvious way.

As for configurations, we require dependencies and children of a component to be disjoint. In order to make some of the forthcoming definitions easier, we allow dependencies to have constraints in the name, origin and version of components. Hence, we may have that the abstract component identifier occurring in a dependency specification is different from the one used in the abstract component of that type. The assumption in this case is that $\forall (t,n,o,v) \in children(cs). \not\exists (t',n',o',v',) \in dependencies(cs). t = t'$.

Now we define configuration specification. In the definition (and in what follows) we will assume a partial order on abstract component identifiers defined as follows: $\leq_{aci}$ be the least partial order on abstract component identifiers such that $(t_1,N_1,O_1,V_1) \leq_{aci} (t_2,N_2,O_2,V_2)$ if and only if $t_1 = t_2, N_1 \subseteq N_2, O_1 \subseteq O_2$ and $V_1 \subseteq V_2$.

\begin{definition}[Configuration specification] A {\em configuration
specification} is a set $CS$ of component specifications such that:
\begin{enumerate}
 \item $\forall cs_1,cs_2 \in CS$ it holds $type (cs_1) \neq type(cs_2)$;
 \item $\forall cs \in CS$ it holds $children(cs) \subseteq acis(CS)$;
 \item $\forall cs \in CS$ it holds $\forall aci \in dependencies(cs). \exists aci' \in acis(CS)$ s.t. $aci \leq_{aci} aci'$;
 \item $\exists ! cs \in CS$ s.t. $\forall cs' \in CS$ it holds $aci(cs)
\not\in children(cs')$;
\item $\forall cs \in C$ either $\not\exists cs'$ s.t. $aci(cs) \in children(cs')$ or $\exists ! cs'$ s.t. $aci(cs) \in children(cs')$
 \item $\forall \component{aci,AD,ACI,[n_1,n_1']} \in CS$ it holds $\boxplus\{[n_2,n_2']
\mid (aci,[n_2,n_2']) \in ACI\} \subseteq [n_1,n_1']$.
\end{enumerate}
\end{definition}

The CSG depicted in \autoref{fig:csg} can be represented as follows. We have the following seven abstract component identifiers:
\begin{gather*}
aci\_Psycho = (Psycho, \{psyN \mid N \in \bbbn\} , \{IMsk\},\bbbn\}\\
aci\_Bin = (Bin, \{binN \mid N \in \bbbn\} , \{IMsk\} , \bbbn\}\\
aci\_PScr = (PScr, Names, Origins, \bbbn\}\\
aci\_CGLib = (CGLib, Names , Origins, \bbbn\}\\
aci\_App = (App, \{psycho\}, \{IMsk\}, \bbbn\}\\
aci\_MLib = (MLib, \{mlib.so\}, \{IMsk\}, \bbbn\}\\
aci\_GLib = (GLib, \{glib.so\}, \{IMsk\}, \bbbn\}
\end{gather*}
that are used in the following representation of the configuration specification:
\begin{align*}
 CS_{Psycho} = \{ \;\;
  & \component{aci\_Psycho,\varnothing,\{(aci\_Bin,[1,1]),(aci\_PScr,[1,\infty]),(aci\_CGLib,[0,\infty])\} , [2,\infty]},\\
  & \component{aci\_Bin,\varnothing,\{(aci\_App,[1,1]),(aci\_MLib,[1,1]),(aci\_GLib,[1,1])\},[3,3]},\\
  & \component{aci\_PScr,\{aci\_PScr,aci\_CGLib\},\varnothing,[0,0]},\\
  & \component{aci\_CGLib,\varnothing,\varnothing,[0,0]},\\
  & \component{aci\_App,\varnothing,\varnothing,[0,0]},\\
  & \component{aci\_MLib,\varnothing,\varnothing,[0,0]},\\
  & \component{aci\_GLib,\varnothing,\varnothing,[0,0]}
\;\; \}
\end{align*}

\subsection{A type discipline for configuration compliance checking}

In this section we formalize the notion of compliance of a configuration with a certain specification.
A configuration specification can be seen as a definition of a {\em type} for configurations, and the compliance of a configuration with respect a certain specification can be seen as a type checking process.
The way in which we perform the type checking is based on a type inference relation and on a subtyping relation. The type inference will allow us to compute a minimal specification satisfied by a given configuration. The subtyping relation will be used to compare the inferred minimal specification with the given one. If the subtyping relation is defined for such two specifications, then the considered configuration is compliant with the given specification.

The type inference relation is based on a notion of unification of configuration specifications.
Given a component identifier $(t,n,o,v)$, let $ci2aci\big((t,n,o,v)\big) =
(t,\{n\},\{o\},\{v\})$ and $ci2aci(CI)$ be its extension to sets of component identifiers defined in the obvious way. Moreover, given two abstract component identifiers
$(t,N_1,O_1,V_1)$ and $(t,N_2,O_2,V_2)$, let $(t,N_1,O_1,V_1) \oplus
(t,N_2,O_2,V_2) = (t,N_1 \cup N_2,O_1 \cup O_2,V_1 \cup V_2)$.

\begin{definition}[Unification] The {\em unification} operation $\Cup$ on
configuration specifications is defined as follows:
\begin{align*}
CS_1 \Cup CS_2 \qqeq & 
	\{cs \mid cs \in CS_1 \wedge \not\exists cs' \in CS_2 .type(cs) =
	type(cs')\}\\
\qqop{\cup} & \{cs \mid cs \in CS_2 \wedge \not\exists cs' \in CS_1 . type(cs) =
	type(cs')\}\\
\qqop{\cup}  &\{\component{aci_1 \oplus aci_2 , AD1 \uplus AD2, ACI_1 \Cup
ACI_2 , [n_1+n_2,n_1'+n_2']} \\
& \mid cs_i = \component{aci_i, ADi, ACI_i, [n_i,n_i']}
\in CS_i \wedge type(cs_1) = type(cs_2)\}
\end{align*}
where the unification of abstract component dependencies is defined as
\begin{align*}
AD_1 \Cup AD_2 \qqeq &
	\{ aci_1 \mid aci_1 \in AD_1 \wedge \not\exists aci_2 \in AD_2. type(aci_1) = type(aci_2)\}\\
\qqop{\cup} &
	\{ aci_2 \mid aci_2 \in AD_2 \wedge \not\exists aci_1 \in AD_1. type(aci_1) = type(aci_2)\}\\
\qqop{\cup} &
	\{ aci_1 \oplus aci_2 \mid aci_i \in AD_i \wedge type(aci_1)=type(aci_2)\}
\end{align*}
and the unification of abstract component children is defined as
\begin{align*}
ACI_1 \Cup ACI_2 \qqeq &
	\{ (aci_1,[0,n_1']) \mid (aci_1,[n_1,n_1']) \in ACI_1 \wedge\\
& \not\exists aci_2,n_2,n_2' . \big((aci_2,[n_2,n_2']) \in ACI_2 \wedge
type(aci_1) = type(aci_2)\big)\}\\
\qqop{\cup} &
	\{ (aci_2,[0,n_2']) \mid (aci_2,[n_2,n_2']) \in ACI_2 \wedge\\
& \not\exists aci_1,n_1,n_1' . \big((aci_1,[n_1,n_1']) \in ACI_1 \wedge
type(aci_1) = type(aci_2)\big)\}\\
\qqop{\cup} &
	\{ (aci_1 \oplus aci_2 , [min(n_1,n_2),max(n1',n2')]) \\ 
& \mid (aci_i,[n_i,n_i']) \in ACI_i \wedge type(aci_1)=type(aci_2)\}
\end{align*}

\end{definition}

The unification operation takes two configuration specifications, which are sets of component specifications, and results into a single set of component specifications.

If a component specification occurs in one configuration and no components of the same type occurs in the other, then the component specification is part of the configuration unification. If components of the same type occur in both configurations, a new component is built in the unification as follows: a) the name of the new component is a merge of the two component names, b) the dependences are the union of the dependences of the two components, c) the subcomponents of the new component result from the unification of the subcomponents, and d) the range of occurrences of the new component in the configuration unification is derived by the intervals of the component specifications.

The unification of abstract component dependencies simply makes the unions of the two sets by merging elements of the same type.

The unification of abstract component children proceeds analogously. When two children have the same type, the interval describing the possible number of components of that type, in the unification, must consider the minimum and the maximum between them.

\begin{proposition}
Given two configuration specifications $CS_1$ and $CS_2$, $CS_1 \Cup CS_2$ is a configuration specification.
\end{proposition}


Now we define a type inference relation that gives the minimal specification satisfied by a given configuration.

\begin{definition}[Type Inference]
The {\em type inference} relation $\vdash C : CS$ is the least relation on
configurations and configuration specifications satisfying the following three rules:
\begin{gather*}
\frac{}
{\vdash \{\component{ci,D,E}\} :
\{\component{ci2aci(ci),ci2aci(D),\varnothing,[1,1]} \}}
\\[0.5\baselineskip]
\frac{ACIs = \{(\oplus CIt, [|CIt|,|CIt|]) \mid CIt \in CI_{/\equiv_t}\}}
 {\vdash \{\component{ci,D,CI}\} :
\{\component{ci2aci(ci),ci2aci(D),ACIs,[|CI|,|CI| ]}\}
}
\\[0.5\baselineskip]
\frac{\vdash C_1 : CS_1 \qquad \vdash C_2 : CS_2}
{\vdash C_1 \cup C_2 : CS_1 \Cup CS_2}
\end{gather*}
where $CI_{/\equiv_t}$ is the quotient set of $CI$ with respect to $\equiv_t$, namely a partition of $CI$ corresponding to the set of all equivalence classes of $\equiv_t$, and
where $\equiv_t$ is the least equivalence on component identifiers such that
$(t_1,n_1,o_1,v_1) \equiv_t (t_2,n_2,o_2,v_2)$ if and only if $t_1 = t_2$.
\end{definition}

The first rule describes the configuration specification of a component without subcomponents. The second rule describes the specification of a component with subcomponents: all the subcomponents with the same type are described by a single component specification.  The third rule simply says that the specification of a union of configurations is the unification of the specifications.

Let us consider again our running example Psycho. It is easy to see that for configurations $C_{psy1}$ and $C_{psy2}$, both $\vdash C_{psy1} : CS_{psy1}$ and $\vdash C_{psy2} : CS_{psy2}$ hold. Let us show only $CS_{psy2}$ (since $CS_{psy1}$ is simpler), that is as follows
\small
\begin{align*}
 & CS_{psy2} = \{ \;\;\\
  & \component{aci\_psy_2,\varnothing,\{(aci\_bin_2,[1,1]),(aci\_def.psc_2 \oplus aci\_my.psc_2,[2,2]),(aci\_julib.so_2,[1,1])\},[4,4]},\\
  & \component{aci\_bin_2,\varnothing,\{(aci\_psycho_2,[1,1]),(aci\_mlib.so_2,[1,1]),(aci\_glib.so_2,[1,1])\},[3,3]},\\
  & \component{aci\_def.psc_2 \oplus aci\_my.psc_2,\{aci\_julib.so_2\},\varnothing,[0,0]},\\
  & \component{aci\_julib.so_2,\varnothing,\varnothing,[0,0]},\\
  & \component{aci\_psycho_2,\varnothing,\varnothing,[0,0]},\\
  & \component{aci\_mlib.so_2,\varnothing,\varnothing,[0,0]},\\
  & \component{aci\_glib.so_2,\varnothing,\varnothing,[0,0]}\\
& \}
\end{align*}
\normalsize
in which $aci\_x$ stands for $ci2aci(ci\_x)$ for any $x$.

\begin{proposition}
Given a configuration $C$, there exists a unique configuration specification $CS$ such that $\vdash C : CS$.
\end{proposition}

Now we define a subtyping relation on configuration specifications.

\begin{definition}[Subtyping]
Let $\leq_{cs}$ be the least partial order on component specifications such that $\component{aci_1,AD_1,ACI_1,[n_1,n_1']} \leq_{cs} \component{aci_2,AD_2,ACI_2,[n_2,n_2']}$ if and only if the following conditions hold:
\begin{itemize}
 \item $aci_1 \leq_{aci} aci_2$;
 \item for all $aci' \in AD_1$ there exists $aci'' \in AD_2$ such that $aci' \leq_{aci} aci''$;
 \item for all $(aci',[n_1'',n_1''']) \in ACI_1$ there exists $(aci'',[n_2'',n_2''']) \in ACI_2$ such that $aci' \leq_{aci} aci''$ and $[n_1'',n_1'''] \subseteq [n_2'',n_2''']$;
 \item $[n_1,n_1'] \subseteq [n_2,n_2']$.
\end{itemize}

The {\em subtyping relation} $\leq_{CS}$ on configuration specifications is the least partial order such that $CS_1 \leq_{CS} CS_2$ holds if and only if for all $cs_1 \in CS_1$ there exists $cs_2 \in CS_2$ such that $cs_1 \leq_{cs} cs_2$.
\end{definition}

The subtyping relation essentialy checks that every component of the smaller configuration specification is contained also in the bigger configuration specification, possibly with weaker constraints.

\begin{definition}[Configuration compliant with a specification]
A configuration $C$ is {\em compliant} with a configuration specification $CS$ if and only if there exists $CS'$ such that $\vdash C : CS'$ and $CS' \leq_{CS} CS$.
\end{definition}

If we consider again our running example Psycho, it is easy to see that, as expected, both configurations $C_{psy1}$ and $C_{psy2}$ are compliant with specification $CS_{Psycho}$. In fact, as already said both $\vdash C_{psy1} : CS_{psy1}$ and $\vdash C_{psy2} : CS_{psy2}$ hold. Moreover, both $CS_{psy1} \leq_{CS} CS_{Psycho}$ and $CS_{psy2} \leq_{CS} CS_{Psycho}$ hold.

A notion of compatibility between configurations can be defined in a way that is similar to the definition of subtyping between specifications.
In this definition we assume backward compatibility for individual components. This means that a component associated with a certain version number is assumed to be compatible with previous versions of the same component.

\begin{definition}[Compatible configurations]
Let $\sqsubseteq_{ci}$ be the least partial order on component identifiers such that $(t_1,n_1,o_1,v_1) \sqsubseteq_{ci} (t_2,n_2,o_2,v_2)$ holds if and only if $t_1=t_2, n_1=n_2, o_1=o_2$ and $v_1 \leq v_2$.

Let $\sqsubseteq_{C}$ be the least partial order on configurations such that for all $c_1 \in C_1$ there exists $c_2 \in C_2$ such that $ci(c_1) \sqsubseteq_{ci} ci(c_2)$.

Given a configuration specification $CS$, a configuration $C_2$ is {\em compatible} with configuration $C_1$ if and only if both $C_1$ and $C_2$ are compliant with $CS$, and $C_1 \sqsubseteq_C C_2$.
\end{definition}

In the Psycho example we have that configuration $C_{psy2}$ as expected is compatible with configuration $C_{psy1}$, since both configurations are compliant with $CS_{Psycho}$ and $C_{psy1} \sqsubseteq C_{psy2}$ holds.

\section{Conclusions}

We have presented a model for software configuration management based on a UML notation and with a formal definition. UML is a well-established language to descrive software system architectures, hence it naturally fits the domain of configuration management. The formal definition gives a solid mathematical foundation to the model and allows us to define a type discipline to check that a configuration is compliant with its specification.

A preliminary version \cite{M06,CM07} of the model has been applied to several case studies of complex software systems which are representative of the scenarios described in \autoref{sec:scenarios}. These case studies include the {\em Spybot} anti-spyware utility (as an example of systems that need very frequent updates), the {\em Cygwin} unix-like environment for Windows (as an example of systems with huge and complex configuration trees) and the {\em Racer} multiplayer drive simulation game (as an example of systems that include lot of user-provided components).

The diffusion of open source software and the consequent possibility for many programmers to participate in the development of software systems makes the management of configurations a challenging issue. Moreover, open source software is often installed and managed by users themselves. In this context the need of flexible and reliable mechanisms for configuration management arises. Current solutions, such as those used to manage linux distributions, are very centralized: users have to follow the evolution of the software components supported by the official repositories. In addition, they generally lack of guarantees of reliability based on some formal theory. We believe that the proposed model fulfills both the need of flexibility and reliability.

Future work might include the definition of a digital format for the representation of specifications and configuration, for instance based on XML/XMI. The development of a managment library supporting operations such as compliance checking, compatibility checking, installation, uninstallation, update, and so on, can be based on such a definition.



\nocite{*}
\bibliographystyle{eceasst}
\bibliography{biblio}

\end{document}